**On the interpretation of the 'extra' variables in the Hurwitz transformation**

Le Van Hoang, Nguyen Thanh Son

Department of Physics, HCMC Education University,
280 An Duong Vuong, District 5, HCM City, Viet Nam,
Email: hoanglv@hcm.fpt.vn

**Abstract**: The Hurwits transformation is generalized by introducing three new variables called "extra". Interpretation of these 'extra' variables allows us to establish relation between isotropic harmonic oscillator and five-dimensional hydrogen-like atom in 'electromagnetic" fields of various configurations. For the Schrödinger equations, the scheme of separation of 'extra' variables is suggested.

FACS: 0365 – w, Fd

## 1. Introduction

By using Hurwitz transformation, which is sometimes called a '5 to 8' generalized Kustaanheimo-Stiefel transformation [1], the relation between the Schrödinger equation for a five-dimensional hydrogen-like atom and that for an isotropic harmonic oscillator in eight-dimensional real space was established long ago [2-3] and has been recently discussed from different points of views (see, for example, refs. [4-9]).

The Hurwitz transformation originally has the form as follows:

$$\begin{aligned}
x_1 &= u_1^2 + u_2^2 + u_3^2 + u_4^2 - u_5^2 - u_6^2 - u_7^2 - u_8^2, \\
x_2 &= 2(u_1 u_5 + u_2 u_6 - u_3 u_7 - u_4 u_8), \\
x_3 &= 2(u_1 u_6 - u_7 u_5 + u_3 u_8 - u_4 u_7), \\
x_4 &= 2(u_1 u_7 + u_2 u_8 + u_3 u_5 + u_4 u_6), \\
x_5 &= 2(u_1 u_8 - u_2 u_7 - u_3 u_6 + u_4 u_5),
\end{aligned} \qquad (1)$$

so any connection between five-dimensional space and the eight-dimensional one:

$$(x_1, x_2, \ldots, x_5) \rightleftarrows (u_1, u_2, \ldots, u_8) \qquad (2)$$

should be companioned with three constraints given for variables $u_1, u_2, \ldots, u_8$. In the paper [4], we have suggested one variant of generalization of transformation (1) by adding three Euler angles $\phi_1, \phi_2, \phi_3$ [10] as new additional variables. The new variable addition does not affect to the main properties of Hurwitz transformation but gives us a way to extract some solutions describing non-trivial topological systems from the wave-functions of a harmonic oscillator. Obviously, in the new form of Hurwitz transformation:

$$(x_1, x_2, \ldots, x_5, \phi_1, \phi_2, \phi_3) \rightleftarrows (u_1, u_2, \ldots, u_8). \qquad (3)$$

the constraints are hidden in the definition of extra variables. Logically, one can try to interpret the additional variables by various ways as it has been done in [11-12] for the case of '3 to 4' KS transformation for obtaining new results from the well considered wave-functions such as of harmonic oscillator. However, only in the paper [6] the additional Euler angles in the Hurwitz transformation were successfully used for obtaining 'new physics', showing that the oscillator is as a hidden non-abelian monopole. For more details one can see also [7], in which this result is well considered from the points of topology theory, formed as Sp(2)-covariant generalization of Hurwitz transformation. The further applications were done in [8-9] for the problem of Coulomb-oscillator duality.

Despite many investigations in this area, some questions are till now open for us: There exists another definition of the 'extra' variables or the given one in [4] is unique? In the case of positive answer, what are the general criteria to define the 'extra' variables? And what the 'new physics' can be obtained from interpretation of these variables besides the five-dimensional Dirac monopole shown in [6] (in [7] the potential obtained is called SU(2)- instanton)? In this paper, we try to answer the above-mentioned questions. Moreover, we will solve the problem of the 'extra' angle separation in the Schrödinger equation of the system obtained when transforming harmonic oscillator to five-dimensional space via the Hurwitz transformation.

## 2. Generalized Hurwitz transformation

Let us consider the transformation as follows:

$$x_\lambda = \xi_s^* (\gamma_\lambda)_{st} \xi_t,$$
$$\phi_k = f_k(\xi), \tag{4}$$

where $\lambda = 1, 2, \ldots, 5$; $k = 1, 2, 3$; the asterisk denotes the complex conjugate operation in four-dimensional complex space with coordinates $\xi_s$ ($s = 1, 2, 3, 4$); repeating indices means summation over them; Dirac matrices $\gamma_\lambda$ are defined as follows:

$$\gamma_k = -i\beta \alpha_k, \quad k = 1, 2, 3$$
$$\gamma_4 = \beta \gamma_1 \gamma_2 \gamma_3, \tag{5}$$
$$\gamma_5 = \beta$$

with $\quad \alpha_k = \begin{pmatrix} 0 & \sigma_k \\ \sigma_k & 0 \end{pmatrix}, \quad \beta = \begin{pmatrix} I & 0 \\ 0 & -I \end{pmatrix};$

$\sigma_k$ ($k = 1, 2, 3$) are the Pauli matrices with properties: $\sigma_k \sigma_j = \delta_{kj} + i\varepsilon_{kjh} \sigma_h.$ \hfill (6)

In transformation (4), $f_k(\xi)$, $k = 1, 2, 3$ are arbitrary functions, which will be defined later. Besides the well-known anti-symmetric properties

$$\gamma_\lambda \gamma_\beta + \gamma_\beta \gamma_\lambda = 2\delta_{\lambda\beta} I,$$

of the Dirac matrices, by using formulae (6), we find the other property, expressed via the following correlation:

$$(\gamma_\lambda)_{st} (\gamma_\lambda)_{uv} = 2\delta_{sv} \delta_{tu} - \delta_{st} \delta_{uv} - 2\tilde{\gamma}_{su} \tilde{\gamma}_{tv}, \tag{7}$$

which is useful for further application. Here we use the antisymmetric matrix $\tilde{\gamma} = i\gamma_1 \gamma_3$ which commutes with all of Dirac matrices. Correlation (7) leads to an important property of the Hurwitz transformation, expressed via the equation:

$$r = \sqrt{x_\lambda x_\lambda} = \xi_s \xi_s^*. \tag{8}$$

In order to receive the suitable form for the differential operators according to variables $\phi_1, \phi_2, \phi_3$ we have some requirements under the functions $f_k(\xi)$, $k = 1, 2, 3$. The first one is to satisfy the following three equations:

$$\xi_s \frac{\partial f_k}{\partial \xi_s} - \xi_s^* \frac{\partial f_k}{\partial \xi_s^*} = -2i\delta_{1k} \; , \tag{9}$$

that leads to an equation:

$$\hat{T}_1 = \frac{1}{2}\left(\xi_s \frac{\partial}{\partial \xi_s} - \xi_s^* \frac{\partial}{\partial \xi_s^*}\right) = -i\frac{\partial}{\partial \phi_1} \; . \tag{10}$$

The second one is that the functions defined as follows

$$\begin{aligned} B_k^+(\phi_1, \phi_2, \phi_3) &= -\frac{1}{2}\tilde{\gamma}_{st}\left(\xi_t \frac{\partial f_k}{\partial \xi_s^*} + \xi_s^* \frac{\partial f_k}{\partial \xi_t}\right), \\ B_k^-(\phi_1, \phi_2, \phi_3) &= \frac{1}{2}i\tilde{\gamma}_{st}\left(\xi_t \frac{\partial f_k}{\partial \xi_s^*} - \xi_s^* \frac{\partial f_k}{\partial \xi_t}\right) \end{aligned} \tag{11}$$

are independent on variables $x_\lambda$. Using this condition (11) and taking into account the fact that matrices $\tilde{\gamma}\gamma_\lambda$ are anti-symmetrical we perform equations:

$$\begin{aligned} \hat{T}_2 &= \frac{1}{2}i\tilde{\gamma}_{st}\left(\xi_t \frac{\partial}{\partial \xi_s^*} + \xi_s^* \frac{\partial}{\partial \xi_t}\right) = -iB_k^+(\phi_1, \phi_2, \phi_3)\frac{\partial}{\partial \phi_k}, \\ \hat{T}_3 &= \frac{1}{2}\tilde{\gamma}_{st}\left(\xi_t \frac{\partial}{\partial \xi_s^*} - \xi_s^* \frac{\partial}{\partial \xi_t}\right) = -iB_k^-(\phi_1, \phi_2, \phi_3)\frac{\partial}{\partial \phi_k}. \end{aligned} \tag{12}$$

Three operators $\hat{T}_1, \hat{T}_2, \hat{T}_3$ generate closed algebra according to the following commutations:

$$\left[\hat{T}_j, \hat{T}_k\right] = i\,\varepsilon_{jkh}\hat{T}_h. \tag{13}$$

From commutations (13), we can interpret extra variables as three Euler angles with the generators of rotor $\hat{T}_1, \hat{T}_2, \hat{T}_3$ correspondingly. Therefore, we can construct three other operators $\hat{Q}_1, \hat{Q}_2, \hat{Q}_3$ with the following properties:

$$\left[\hat{Q}_j, \hat{Q}_k\right] = -i\, \varepsilon_{jkh}\, \hat{Q}_h, \qquad \left[\hat{T}_j, \hat{Q}_k\right] = 0, \qquad \hat{T}_k \hat{T}_k = \hat{Q}_j \hat{Q}_j = \hat{Q}^2. \tag{14}$$

For the physical meaning of the operators $\hat{Q}_1, \hat{Q}_2, \hat{Q}_3$ concerning to the Euler angles, one can see, for example, in the book [10]. Hereby we use the fact that operators $\hat{Q}_1, \hat{Q}_2, \hat{Q}_3$ coincide with $\hat{T}_1, \hat{T}_2, \hat{T}_3$ after the change:

$$\phi_1 \rightleftarrows \phi_2, \quad \phi_3 \to -\phi_3 \tag{15}$$

to have:

$$\hat{Q}_1 = -i\frac{\partial}{\partial \phi_2},$$

$$\hat{Q}_2 = -i\, B_2^{+}(\phi_2, \phi_1, -\phi_3)\frac{\partial}{\partial \phi_1} - i\, B_1^{+}(\phi_2, \phi_3, -\phi_3)\frac{\partial}{\partial \phi_2} - i\, B_3^{+}(\phi_2, \phi_1, -\phi_3)\frac{\partial}{\partial \phi_3}, \tag{16}$$

$$\hat{Q}_3 = -i\, B_2^{-}(\phi_2, \phi_1, -\phi_3)\frac{\partial}{\partial \phi_1} - i\, B_1^{-}(\phi_2, \phi_1, -\phi_3)\frac{\partial}{\partial \phi_2} - i\, B_3^{-}(\phi_2, \phi_1, -\phi_3)\frac{\partial}{\partial \phi_3}.$$

Now we consider the momentum operator in $\xi$- space. After some differential conversions with using transformation (4), we find the correlation:

$$\frac{1}{2}(\gamma_\lambda)_{st}\left(\xi_t \frac{\partial}{\partial \xi_s} + \xi_s^* \frac{\partial}{\partial \xi_t^*}\right) = r\left(\frac{\partial}{\partial x_\lambda} + \tilde{A}_{\lambda k}(\mathbf{r}, \phi)\frac{\partial}{\partial \phi_k}\right). \tag{17}$$

Here in (17):

$$\tilde{A}_{\lambda k}(\mathbf{r}, \phi) = \frac{1}{2r}(\gamma_\lambda)_{st}\left(\xi_t \frac{\partial f_k}{\partial \xi_s} + \xi_s^* \frac{\partial f_k}{\partial \xi_t^*}\right). \tag{18}$$

The form of equation (17) prompts us to define the momentum operator as follows:

$$\hat{p}_\lambda = -i\frac{\partial}{\partial x_\lambda} - i\,\tilde{A}_{\lambda k}(\mathbf{r},\phi)\frac{\partial}{\partial \phi_k}.$$

For the angle separation procedure given below in Section 5, we will rewrite it into a more convenient form. Indeed, with the use of (16) we have:

$$\hat{p}_\lambda = -i\frac{\partial}{\partial x_\lambda} + A_{\lambda k}\,\hat{Q}_k$$

with the functions $A_{\lambda k}$ defined via $\tilde{A}_{\lambda k}(\mathbf{r},\phi)$ as follows:

$$\begin{aligned}
A_{\lambda 1} &= \tilde{A}_{\lambda 2} - \frac{B_3^+ B_1^- - B_3^- B_1^+}{B_3^+ B_2^- - B_2^+ B_3^-}\tilde{A}_{\lambda 1} + \frac{B_2^+ B_1^- - B_2^- B_1^+}{B_3^+ B_2^- - B_2^+ B_3^-}\tilde{A}_{\lambda 3}\,. \\
A_{\lambda 2} &= -\frac{B_3^-}{B_3^+ B_2^- - B_2^+ B_3^-}\tilde{A}_{\lambda 1} + \frac{B_2^-}{B_3^+ B_2^- - B_2^+ B_3^-}\tilde{A}_{\lambda 3}\,, \\
A_{\lambda 3} &= \frac{B_3^+}{B_3^+ B_2^- - B_2^+ B_3^-}\tilde{A}_{\lambda 1} - \frac{B_2^+}{B_3^+ B_2^- - B_2^+ B_3^-}\tilde{A}_{\lambda 3}\,,
\end{aligned} \quad (19)$$

where all the functions $B_k^\pm(\phi_2,\phi_1,-\phi_3)$ have arguments corresponding to the changes (15). The third requirement for the functions $f_k(\xi)$, $k=1,2,3$ is that they should be chosen in the way that the functions $A_{\lambda k}$ depend on coordinates $\mathbf{r}$ only, i.e. independent on variables $\phi_1,\phi_2,\phi_3$. Thus, the momentum operator can be written in the form:

$$\hat{p}_\lambda = -i\frac{\partial}{\partial x_\lambda} + A_{\lambda k}(\mathbf{r})\,\hat{Q}_k = -\frac{i}{2\xi_u \xi_u^*}(\gamma_\lambda)_{st}\left(\xi_t \frac{\partial}{\partial \xi_s} + \xi_s^* \frac{\partial}{\partial \xi_t^*}\right). \quad (20)$$

Consequently, we have generalized the Hurwits transformation. In this generalization, we introduce three 'extra' variables and give three criteria to define them. In next sections, we will show that the suitable interpretation of these extra variables leads to new interesting results. The interpretation is analogous to that we have done in [11-12] for the extra variable in generalized KS-transformation. The difference is that in our case there are three extra variables despite one extra variable in the case of '3 to 4' KS-transformation [13].

## 3. Relation between harmonic oscillator and hydrogen-like atom in magnetic field of various configurations

Let us consider the Schrödinger equation for an isotropic harmonic oscillator in $\xi$- space:

$$\hat{H}\psi(\xi) = Z\psi(\xi), \qquad (21)$$

$$\hat{H} = -\frac{1}{2}\frac{\partial^2}{\partial \xi_s \partial \xi_s^*} + \frac{1}{2}\omega^2 \xi_s \xi_s^*, \qquad (22)$$

where frequency $\omega$ is a real positive number. In order to transform equation (21)-(22) into the real $\mathbf{r}$ - space, we firstly calculate the term $\hat{\mathbf{p}}^2 = \hat{p}_\lambda \hat{p}_\lambda$. By using correlation (20) and formula (7), we obtain:

$$r\left(-i\frac{\partial}{\partial x_\lambda} + A_{\lambda k}(\mathbf{r})\hat{Q}_k\right)^2 + \frac{1}{2r}\hat{Q}^2 = -\frac{\partial^2}{\partial \xi_s \partial \xi_s^*}. \qquad (23)$$

By substituting (8) and (23) into (20)-(21), we find:

$$\left\{\frac{1}{2}\left(-i\frac{\partial}{\partial x_\lambda} + A_{\lambda k}(\mathbf{r})\hat{Q}_k\right)^2 + \frac{1}{2r^2}\hat{Q}^2 - \frac{Z}{r}\right\}\psi(\mathbf{r},\phi) = E\psi(\mathbf{r},\phi), \qquad (24)$$

where $E = -\frac{1}{2}\omega^2$ is considered as energy of the system. For the class of wave-functions, which are independent on the 'extra' variables $\phi_1, \phi_2, \phi_3$, i.e.

$$\hat{Q}_1\psi = \hat{Q}_2\psi = \hat{Q}_3\psi = 0, \qquad (25)$$

the equation (24) describes the five-dimensional hydrogen-like atom. It means that from solutions of equation (21)-(22) we can extract wave-functions for five-dimensional hydrogen-like atom by using conditions (25). However, we should pay attention to the fact that if in equation (21)-(22) the value Z is an eigen-value, then in equation (24) it just plays the role of parameter. In general, equation (21)-(22) is equivalent to equation (24). In other words, they are dual to each other. But the last describes the motion of 'isospin' particle in a five-dimensional Coulomb potential and interacting with 'electromagnetic' vector potential, the configuration of which depends upon the definition of extra variables $\phi_1, \phi_2, \phi_3$. In section 6, we will show how to separate 'extra' variables from equation (24). In the next section, we just interpret these variables to receive the various forms of 'electromagnetic' field.

## 4. Interpretation of 'extra' variables

Let us now define the variables $\phi_1, \phi_2, \phi_3$ by choosing the functions $f_1, f_2, f_3$ satisfying the conditions mentioned in section 2. For this purpose we solve the equations (9), obtained from the first demand. The general solutions can be found in the form:

$$\begin{aligned}
f_1(\xi) &= \arg \xi_1 + \arg \xi_2 + F_1(\xi_j \xi_k^*), \\
f_2(\xi) &= \arg \xi_1 - \arg \xi_2 + F_2(\xi_j \xi_k^*), \\
f_3(\xi) &= \arctan \frac{2\sqrt{\xi_1 \xi_1^* \xi_2 \xi_2^*}}{\xi_1 \xi_1^* - \xi_2 \xi_2^*} + F_3(\xi_j \xi_k^*),
\end{aligned} \qquad (26)$$

where $F_1(\xi_j \xi_k^*), F_2(\xi_j \xi_k^*), F_3(\xi_j \xi_k^*)$ are arbitrary functions.

**Case A:** The case of $F_1(\xi_j \xi_k^*) = F_2(\xi_j \xi_k^*) = F_3(\xi_j \xi_k^*) = 0$ leads to our previous result published in [4] and considered in [6]:

$$\begin{aligned}
\phi_1 &= \arg \xi_1 + \arg \xi_2, \\
\phi_2 &= \arg \xi_1 - \arg \xi_2, \\
\phi_3 &= \arctan \frac{2\sqrt{\xi_1 \xi_1^* \xi_2 \xi_2^*}}{\xi_1 \xi_1^* - \xi_2 \xi_2^*}.
\end{aligned} \qquad (27)$$

By substituting (27) into expressions (11)-(16) we acquire an obvious form of $\hat{Q}_1, \hat{Q}_2, \hat{Q}_3$ as follows:

$$\hat{Q}_1 = -i\frac{\partial}{\partial \phi_2},$$

$$\hat{Q}_2 = -i\frac{\cos\phi_2}{\sin\phi_3}\frac{\partial}{\partial \phi_1} + i\frac{\cos\phi_2}{\tan\phi_3}\frac{\partial}{\partial \phi_2} + i\sin\phi_2\frac{\partial}{\partial \phi_3}, \quad (28)$$

$$\hat{Q}_3 = -i\frac{\sin\phi_2}{\sin\phi_3}\frac{\partial}{\partial \phi_1} + i\frac{\sin\phi_2}{\tan\phi_3}\frac{\partial}{\partial \phi_2} - i\cos\phi_2\frac{\partial}{\partial \phi_3}.$$

Evidently, operators (28) are the generators corresponding to the rotor transformation by three Euler angles (27), $0 \leq \phi_1 \leq 2\pi, 0 \leq \phi_2 \leq 2\pi, 0 \leq \phi_3 \leq \pi$ [10]. This fact will be used in next section to separate the extra variable dependence from equation (24). Putting (27) into (11) and then into (19) we obtain:

$$A_{\lambda 1} = \frac{1}{r(r+x_5)}(x_2, -x_1, -x_4, x_3, 0),$$

$$A_{\lambda 2} = \frac{1}{r(r+x_5)}(-x_4, x_3, -x_2, x_1, 0), \quad (29)$$

$$A_{\lambda 3} = \frac{1}{r(r+x_5)}(x_3, x_4, -x_1, -x_2, 0).$$

It is easy to verify that triplet of vectors (29) satisfies the following properties:

$$A_{\lambda k} A_{\lambda j} = \frac{r-x_5}{r^2(r+x_5)}\delta_{kj}, \quad x_\lambda A_{\lambda k} = 0$$

and has singularity line along the negative part of axis $x_5$. Each term of (29) concerns to the five-dimensional vector potential of Dirac monopole [6] (it is also called SU(2) – instanton [7]).

**Case B.** In the case of

$$\phi_1 = \arg\xi_3 + \arg\xi_4,$$
$$\phi_2 = \arg\xi_3 - \arg\xi_4, \quad (30)$$
$$\phi_3 = \arctan\frac{2\sqrt{\xi_3\xi_3^*\xi_4\xi_4^*}}{\xi_3\xi_3^* - \xi_4\xi_4^*},$$

we obtain the operators $\hat{Q}_1, \hat{Q}_2, \hat{Q}_3$ which coincide with the case A, i.e. have the form (27). However, the vector potentials are different from (28) and have the form:

$$A_{\lambda 1} = \frac{1}{r(r-x_5)}(-x_2, x_1, -x_4, -x_3, 0),$$

$$A_{\lambda 2} = \frac{1}{r(r-x_5)}(-x_4, -x_3, x_2, -x_1, 0), \qquad (31)$$

$$A_{\lambda 3} = \frac{1}{r(r-x_5)}(-x_3, x_4, x_1, x_2, 0).$$

We can see that the triplet of vectors (31) is indeed the same as (29). The difference is only in the singularity line, which is along the positive part of axis $x_5$ in this case. Thus, the angle definition (30) changes the direction of Dirac monopole only. In principle, we can choose the angles (30) in the way that the Dirac monopole has any given direction. Moreover, the combination of A and B leads to the singularity line of potential vectors of the whole axis $x_5$, which has extremely special property similar to the Aharonov-Bohm potential in the three-dimensional space [14-15]. This potential has the topology property different from the Dirac monopole potential and thus requires the particular investigation in another work.

## 5. Angle Separation in the Equation

For the case when variables $\phi_1, \phi_2, \phi_3$ are defined as Euler angles like the cases A, B, considered in section 4, we now show how to separate these angles from equation (24). Noting that the Hamiltonian (22) commutes with all of operators $\hat{Q}_1, \hat{Q}_2, \hat{Q}_3$ and $\hat{T}_1, \hat{T}_2, \hat{T}_3$, we can build the angular part of wave-functions by demanding them to belong to eigen-functions of the operators $\hat{Q}^2, \hat{Q}_1, \hat{T}_1$:

$$\hat{Q}^2 \varphi_{q,p}^J(\phi) = J(J+1) \varphi_{q,p}^J(\phi),$$

$$\hat{Q}_1 \varphi_{q,p}^J(\phi) = q \varphi_{q,p}^J(\phi), \qquad (32)$$

$$\hat{T}_1 \varphi_{q,p}^J(\phi) = p \varphi_{q,p}^J(\phi),$$

with $J = 0, 1, 2, \ldots$ ; $q, p = -J, -J+1, \ldots, J-1, J$ .

Here the operators $\hat{T}_1, \hat{T}_2, \hat{T}_3$ are defined as follows:

$$\hat{T}_1 = -i\frac{\partial}{\partial \phi_1},$$
$$\hat{T}_2 = -i\frac{\sin\phi_1}{\tan\phi_3}\frac{\partial}{\partial \phi_1} + i\frac{\sin\phi_1}{\sin\phi_3}\frac{\partial}{\partial \phi_2} + i\cos\phi_1\frac{\partial}{\partial \phi_3}, \qquad (33)$$
$$\hat{T}_3 = -i\frac{\cos\phi_1}{\tan\phi_3}\frac{\partial}{\partial \phi_1} + i\frac{\cos\phi_1}{\sin\phi_3}\frac{\partial}{\partial \phi_2} + i\sin\phi_1\frac{\partial}{\partial \phi_3}.$$

which have the properties (13)-(14), mentioned in Section 2.

For the physical meaning of the operators $\hat{T}_1, \hat{T}_2, \hat{T}_3$ and for the obvious form of the functions $\varphi_{q,p}^J(\phi)$, one can see in [10]. Here we just want to write some formulae to be used later on:

$$\hat{Q}_\pm \varphi_{q,p}^J(\phi) = \sqrt{(J \mp q)(J \pm q + 1)}\, \varphi_{q \pm 1, p}^J(\phi) , \qquad (34)$$

where $\hat{Q}_\pm = \hat{Q}_2 \pm i\hat{Q}_3$. For the angle separation process, we construct the wave-functions of equation (24) in the form:

$$\psi_{Jp}(\mathbf{r}, \phi) = \Psi_{Jp}(\mathbf{r}) G_{Jp}(\mathbf{r}, \phi), \qquad (35)$$

with 
$$G_{Jp}(\mathbf{r}, \phi) = \sum_{q=-J}^{J} g_q(\mathbf{r}) \varphi_{pq}^J(\phi). \qquad (36)$$

Using the notation:

$$\hat{A}_\lambda = A_{\lambda k}(\mathbf{r})\hat{Q}_k = A_{\lambda 1}(\mathbf{r})\hat{Q}_1 + A_{\lambda +}(\mathbf{r})\hat{Q}_+ + A_{\lambda -}(\mathbf{r})\hat{Q}_-, \qquad (37)$$

where $A_{\lambda +}(\mathbf{r}) = \frac{1}{2}(A_{\lambda 2}(\mathbf{r}) - i A_{\lambda 3}(\mathbf{r}))$, $A_{\lambda -}(\mathbf{r}) = \frac{1}{2}(A_{\lambda 2}(\mathbf{r}) + i A_{\lambda 3}(\mathbf{r}))$

we choose the coefficients $g_q(\mathbf{r})$ in the way that the angular part of wave-function (39) satisfies the equation:

$$\hat{A}_\lambda G_{Jp}(\mathbf{r},\phi) = a_\lambda(\mathbf{r}) G_{Jp}(\mathbf{r},\phi). \tag{38}$$

Besides, from (32) we have the two other equations for the angular part too:

$$\hat{Q}^2 G_{Jp}(\mathbf{r},\phi) = J(J+1)\, G_{Jp}(\mathbf{r},\phi), \tag{39}$$

$$\hat{T}_1\, G_{Jp}(\mathbf{r},\phi) = p\, G_{Jp}(\mathbf{r},\phi). \tag{40}$$

Now let us resolve equation (38) by substituting (36), (37) into it and then using formula (34). As a result, for each term $\lambda = 1, 2, ..., 5$ we obtain a system of $2J+1$ linear uniform equations:

$$\sqrt{(J-q+1)(J+q)}\, A_{\lambda+}(\mathbf{r}) g_{q-1}(\mathbf{r}) - \left[a_\lambda(\mathbf{r}) - q A_{\lambda 1}(\mathbf{r})\right] g_q(\mathbf{r})$$

$$+ \sqrt{(J+q+1)(J-q)}\, A_{\lambda-}(\mathbf{r}) g_{q+1}(\mathbf{r}) = 0, \tag{41}$$

for $q = -J, -J+1, -J+2, ..., J-1, J$. Since system of equations (44) is uniform, it has nontrivial solution only in the case when the determinant of the following matrix:

$$h_\lambda = \begin{pmatrix} -a_\lambda - J A_{\lambda 1} & \sqrt{2J} A_{\lambda-} & 0 & \ldots & 0 \\ \sqrt{2J} A_{\lambda+} & -a_\lambda - (J-1) A_{\lambda 1} & \sqrt{2(2J-1)} A_{\lambda-} & \ldots & 0 \\ 0 & \sqrt{2(2J-1)} A_{\lambda+} & -a_\lambda - (J-2) A_{\lambda 1} & \ldots & 0 \\ \ldots & \ldots & \ldots & \ldots & \sqrt{2J} A_{\lambda-} \\ 0 & 0 & 0 & \sqrt{2J} A_{\lambda+} & -a_\lambda + J A_{\lambda 1} \end{pmatrix}$$

is vanished. Noticing that only the following elements of matrix $h_\lambda$ are different from zero

$$(h_\lambda)_{k,k-1} = \sqrt{(2J+2-k)(k-1)}\, A_{\lambda+}(\mathbf{r}),$$

On the interpretation of 'extra' variables in the Hurwitz transformation

$$(h_\lambda)_{kk} = -a_\lambda(\mathbf{r}) - (J+1-k)A_{\lambda 1}(\mathbf{r}), \tag{42}$$

$$(h_\lambda)_{k,k+1} = \sqrt{k(2J+1-k)}A_{\lambda-}(\mathbf{r}) \ , k=1,2,\ldots,2J+1,$$

the condition $\det(h_\lambda) = 0$ leads to an algebraic equation with order of $2J+1$ for the root $a_\lambda(\mathbf{r})$. Putting the found $a_\lambda(\mathbf{r})$ into equations (41) we can find $2J+1$ coefficients $g_q(\mathbf{r})$ for each term $\lambda$. Thus, we have found the angular part of wave-functions $G_{Jp}(\mathbf{r},\phi)$. Now let us put the wave functions (36) into equation (24), taking into account the equations (38), (39), (40), we obtain the equation for the motion in 'physical' $\mathbf{r}$ – space:

$$\left\{\frac{1}{2}\left(-i\frac{\partial}{\partial x_\lambda} - a_\lambda(\mathbf{r})\right)^2 + \frac{J(J+1)}{2r^2} - \frac{Z}{r}\right\}\Psi_{Jp}(\mathbf{r}) = E\Psi_{Jp}(\mathbf{r}). \tag{43}$$

Consequently, we have suggested one variant to construct the relation between an isotropic harmonic oscillator in four-dimensional complex space and a five-dimensional hydrogen-like atom in the electromagnetic field $\left(\mathbf{a}(\mathbf{r}), \frac{J(J+1)}{2r^2}\right)$. For concrete application, we consider the case of $J=1$. The matrix $h_\lambda$ becomes

$$h_\lambda = \begin{pmatrix} -a_\lambda - A_{\lambda 1} & \sqrt{2}A_{\lambda-} & 0 \\ \sqrt{2}A_{\lambda+} & -a_\lambda & \sqrt{2}A_{\lambda-} \\ 0 & \sqrt{2}A_{\lambda+} & -a_\lambda + A_{\lambda 1} \end{pmatrix},$$

and the condition $\det(h_\lambda) = 0$ leads to the following equation:

$$a_\lambda(\mathbf{r})\left(a_\lambda^2(\mathbf{r}) - A_{\lambda 1}^2(\mathbf{r}) - A_{\lambda 2}^2(\mathbf{r}) - A_{\lambda 3}^2(\mathbf{r})\right) = 0. \tag{44}$$

Thus, besides the zero solution we have:

$$a_\lambda(\mathbf{r}) = \pm\sqrt{A_{\lambda 1}^2(\mathbf{r}) + A_{\lambda 2}^2(\mathbf{r}) + A_{\lambda 3}^2(\mathbf{r})}\ .$$

For the case A of section 4, after substituting the concrete formulae (29) we obtain:

$$a_1(\mathbf{r}) = -\frac{\sqrt{x_2^2 + x_3^2 + x_4^2}}{r(r+x_5)}, \quad a_2(\mathbf{r}) = \frac{\sqrt{x_1^2 + x_3^2 + x_4^2}}{r(r+x_5)}, \quad a_3(\mathbf{r}) = -\frac{\sqrt{x_1^2 + x_2^2 + x_4^2}}{r(r+x_5)},$$
$$a_4(\mathbf{r}) = \frac{\sqrt{x_1^2 + x_2^2 + x_3^2}}{r(r+x_5)}, \quad a_5(\mathbf{r}) = 0,$$
(45)

here the signs are chosen by the way that (45) is the vector generalized from the Dirac Monopole potential for the case of three-dimensional space.

## 6. Summary and Discussion

Thus, in order to answer the questions given in Introduction we have received the general criteria to define the extra variables adding to the Hurwitz transformation. The generalized Hurwitz transformation built with the extra variables is suitable to establish a connection between an isotropic harmonic oscillator in four-dimensional complex space and five-dimensional hydrogenic oscillator in 'electromagnetic' field of various configurations. We have showed that the form and properties of 'electromagnetic' field are dependent on the definition of the extra variables. In other words, the interpretation of these variables leads to a variety of potentials including the one of five-dimensional Dirac monopole. The general formulae obtained allow us to choose the extra variable in the way to achieve the potential, which has the topology basically different from the one of Dirac monopole. Consequently, it is required that this case be considered carefully and with more details in other work.

　　　The equation obtained from Schrödinger equation for harmonic oscillator principally describes the isospin particle moving in the Coulomb potential and interacting with the vector potential depended on the extra variable definition. Here, the isospin or an internal structure of particle is described by three generators of rotor operators of extra angles, introduced into the generalized Hurwitz transformation. Thus, extra variables are related to describing internal structure of particle. We have suggested scheme of separation of these variables. In other words, we have successfully extracted the isospin dependence from the equation and, as a result, received the equation described motion of a 'bare' particle without isospin.